# First Field Trial of LLM-Powered AI Agent for Lifecycle Management of Autonomous Driving Optical Networks

Xiaomin Liu, Qizhi Qiu, Yihao Zhang, Yuming Cheng, Lilin Yi, Weisheng Hu, and Qunbi Zhuge[*]

State Key Laboratory of Advanced Optical Communication Systems and Networks, Department of Electronic Engineering, Shanghai Jiao Tong University, Shanghai, China, [*]qunbi.zhuge@sjtu.edu.cn

**Abstract** *We design and demonstrate the first field trial of LLM-powered AI Agent for ADON. Three operation modes of the Agent are proposed for network lifecycle management. The Agent efficiently processes wavelength add/drop and soft/hard failures, and achieves comparable performance to human-designed algorithms for power optimization.* ©2024 The Author(s)

## Introduction

Autonomous driving optical networks (ADON) are being developed to improve network efficiency, reliability and performance [1-2]. However, it is extremely challenging to design comprehensive solutions in the conventional manner for all possible events throughout a lifecycle of an optical network. Recently, the remarkable advances in large language models (LLMs) have catalysed the emergence of LLM-powered AI Agents [3-4]. Such an Agent is an intelligent entity that leverages LLM as its brain to generate strategies and employ tools to interact with the physical environment and tackle complex tasks. For optical networks, the Agent holds the potentials to realize a full ADON. To date, an AI Agent for the operations of the entire ADON lifecycle has yet to be investigated and developed.

In this paper, we design and demonstrate the first field trial of an LLM-powered AI Agent to automate lifecycle-level ADON operations. We develop three distinct operation modes of the Agent, tailored to leverage current LLMs' capabilities in addressing diverse optical network events. In the field-trial testbed, typical events during a network lifecycle are emulated and autonomously handled by the Agent from the beginning of life (BoL) to the end of life (EoL), covering service establishment, wavelength add/drop, and hard/soft failures. During these events, the Agent autonomously conducts optical amplifier (OA) optimization to achieve a near-optimal Q-factor, constructs a digital twin (DT) model of the link with a high accuracy, and successfully localize and recover failures including fiber cut and link aging within tens of seconds. Moreover, we compare the capabilities of various LLMs in directly optimizing OAs, further demonstrating the potential of LLM in tackling complex tasks for ADON.

## LLM-powered AI Agent for ADON

Fig. 1 shows the proposed AI Agent to enable autonomous network operations throughout the lifecycle. The central unit of the AI Agent is a Planner, which performs strategy and workflow generation, information analysis, and decision-making. Moreover, the Agent comprises four modules: 1) Knowledge Retrieval (K. R.) to provide domain knowledge; 2) Monitoring & Analytics (M. A.) to collect data and conduct data analysis; 3) DT Construction to synchronize digital models with the real-time network; 4) Management & Control (M. C.) to conduct network operations such as optical power optimization.

We define three operation modes of the AI Agent based on the LLM's capability for different tasks. 1) **LLM-native mode**: the Agent exclusively relies on LLM for planning, reasoning, and execution without any human-designed algorithms; 2) **LLM-centric mode**: the Agent employs LLM for planning and reasoning while leveraging some human-designed algorithms; 3) **Rule-centric mode**: the LLM-powered Agent guides the Planner to retrieve and execute pre-defined workflows with existing tools/algorithms. Notably, these modes are all autonomous without human intervention, and LLM can be designed to perform mode selection for a given task.

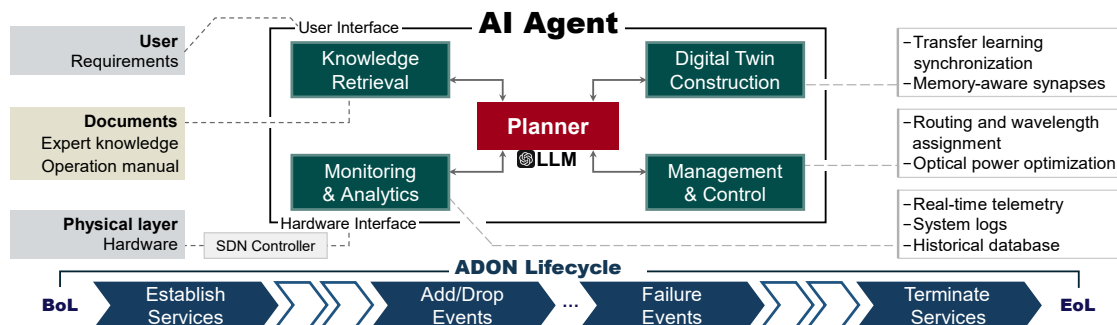

**Fig. 1.** Architecture of the proposed LLM-powered AI Agent for ADON.

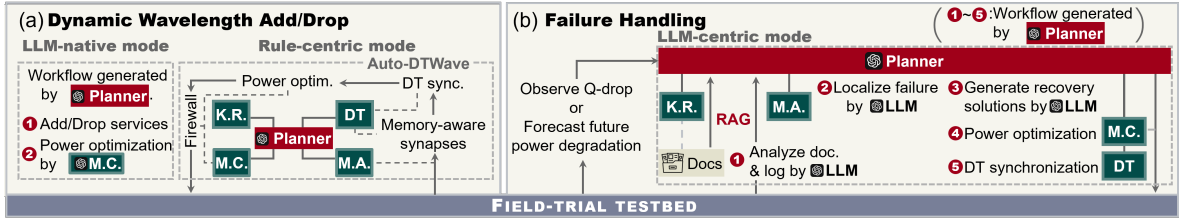

**Fig. 2.** The flow diagrams of the Agent-enabled (a) dynamic wavelength add/drop and (b) failure handling.

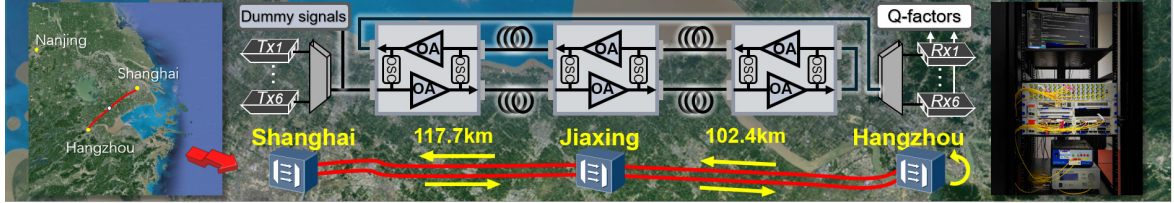

**Fig. 3.** Field-trial testbed.

In the field trial, we demonstrate the applications of all the three modes, and the flow diagrams are shown in Fig. 2. For dynamic wavelength add/drop operations, we illustrate both the LLM-native mode and Rule-centric mode as per Fig. 2(a). For the former, the Agent adds or drops wavelengths first and then performs power optimization. We propose to use the LLM to directly optimize the OA gains based on a ReAct framework with an iterative thought-action-observation loop [5]. For the latter, the pre-defined workflow and algorithm are based on Auto-DTWave [6], which can simultaneously optimize power and synchronize DT.

Fig. 2(b) illustrates how the Agent handles failures in the LLM-centric mode. The process is initiated upon the detection of a Q-factor drop or the forecast of power degradation by the M. A. module. The LLM within the Planner then formulates a workflow, which involves the LLM analysing documents and logs via Retrieval Augmented Generation (RAG) [7] ①, localizing the failure ②, and generating recovery solutions ③. Notably, steps ② and ③ rely solely on the LLM without external algorithms. Then, the Planner autonomously executes the recovery process by conducting power optimization ④ while synchronizing the DT accordingly ⑤.

**Field-trial Testbed and Agent Implementation**
Fig. 3 plots our field-deployed testbed with commercial equipment from Shanghai via Jiaxing to Hangzhou in China. It consists of a 440-km loop link with four spans of G.652.D fibers and six OAs. Six transponders operating at 63.9 GBaud and 200 Gbps are adopted. We emulate a transmission scenario with up to 30 wavelengths in a 75-GHz fixed grid. The wavelengths are added or dropped in batches. Each batch contains one real signal and four dummy signals.

A software-defined network (SDN) controller embedded with the Agent is developed. The Agent is built based on LangChain using the plan-and-execute framework [8]. The control commands are launched from the Agent to the hardware through the NETCONF protocol using YANG models. Our testbed supports millisecond-level telemetry data collection, which significantly enhances the capability of the Agent. These telemetry data are transmitted through the optical supervisory channel (OSC) to the SDN server. Plentiful operation and device logs are stored in the SDN server for the Agent to generate more effective strategies.

**Field-trial Results**
Typical events throughout a network lifecycle were emulated as shown in Fig. 4(b), and we adopted GPT-4o in the Agent to address them [9]. At the BoL, 4 batches of wavelengths (20 wav.) were established. During our experiment, an actual fiber cut truly occurred between Jiaxing and Shanghai, and this was included as a hard failure event. After the recovery of it, service add/drop events were generated by changing the link load from 20 to 30, 25, and 15 wavelengths. Then, to emulate progressive link aging, we employed variable optical attenuators to gradually increase fiber attenuation. Finally, new services were added to increase the link load to 30 wavelengths.

*Rule-centric mode performance.* We first employ the Agent with the Rule-centric mode for the wavelength add/drop events. The Agent autonomously executes the workflow of Auto-DTWave [6] for DT synchronization and power optimization by adjusting the gain and tilt of each OA. Fig. 4(a) shows the Q-factors' changes during the emulated lifecycle events (Fig. 4(b)). The Agent's converged Q-factor deviates <0.3-dB on average from the brute-force optimum. To test the DT's accuracy, three sets of 300 data each are collected, with different loadings and OA configurations at three stages: 1) before the fiber cut, 2) after the fiber cut, and 3) after the link aging

Version submitted to ECOC PDP 2024 on September 6th

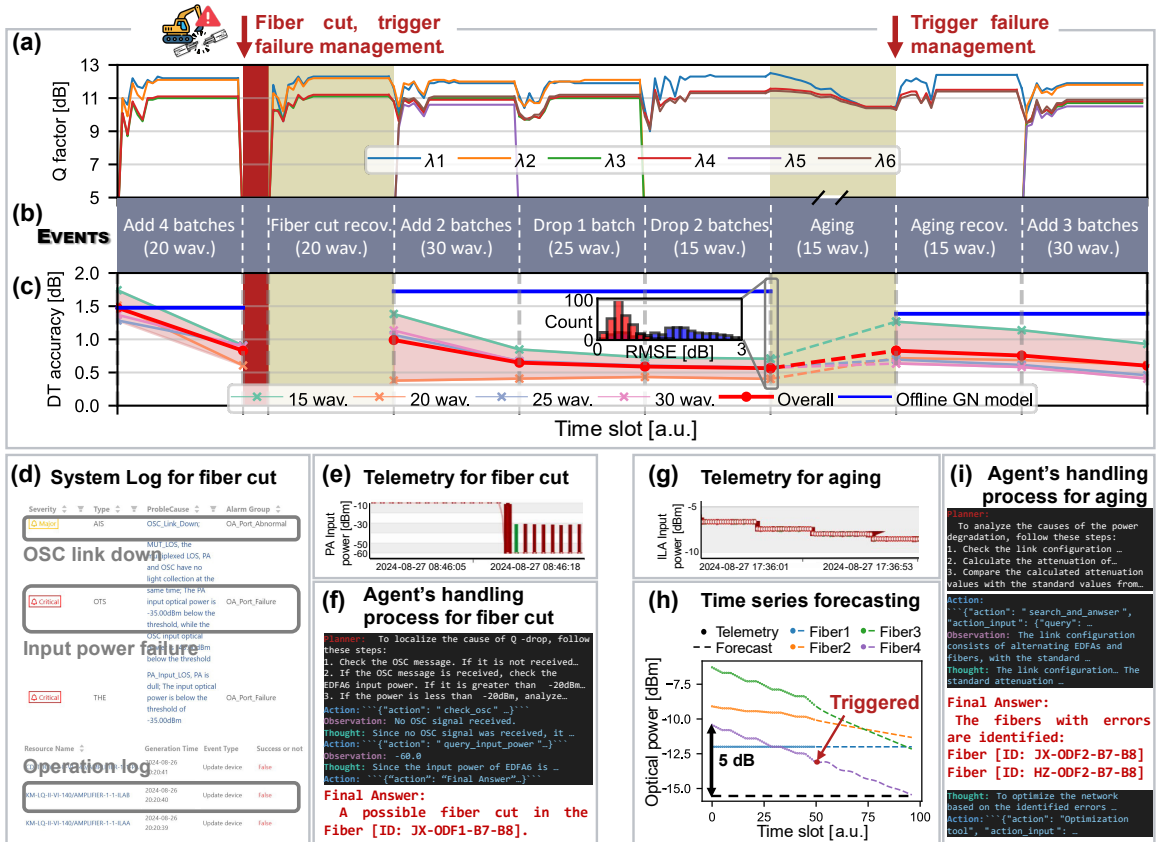

**Fig. 4.** (a) Q-factor variations during service add/drop, failure occurrence, and failure recovery. (b) The investigated events during the emulated lifecycle. (c) The corresponding DT estimation error over various independent test sets. (d-f) The system log, real-time telemetry, and AI Agent's outputs during the real fiber cut event. (g-i) The real-time telemetry, time series forecasting, and AI Agent's outputs during the link aging event.

event. In Fig. 4(c), compared to the offline Gaussian noise model [10] with known parameters, the DT constructed by the Agent reduces the root mean square error from 1.72 dB to 0.56 dB.

*LLM-centric mode performance.* Fig. 4(d)-(f) illustrate the Agent's fiber cut management in the LLM-centric mode. Based on the LLM aided by operation manuals, the Agent autonomously generates a workflow, and then analyses system logs, OSC messages, and telemetry data. Finally, it successfully identifies the fiber cut between Jiaxing and Shanghai in 13 seconds. Fig. 4(g)-(i) illustrate the AI Agent's performance in dealing with link aging. As shown in Fig. 4(h), the M. A. module triggers the failure handling upon predicting a potential 5-dB power loss. The Planner then autonomously generates a workflow using the LLM. Accordingly, the Agent retrieves datasheets and queries optical powers to calculate the attenuation of each span within a minute. Then, it reconfigures the OAs to recover the transmission performance degradation. The above results show the strong reasoning capability of the LLM as it can process diverse failure-related information across multiple sources and formats.

*LLM-native mode performance.* Finally, we evaluate the capabilities of LLMs in optimizing the OA gains in the LLM-native mode. As shown in Fig. 5, the 7 LLMs developed by 4 companies show very different performance. Impressively, 3 of them can achieve results comparable to the Bayesian optimization (BO) approach [11], demonstrating great potential to address intractable tasks that lack solutions today, especially considering they are still rapidly evolving.

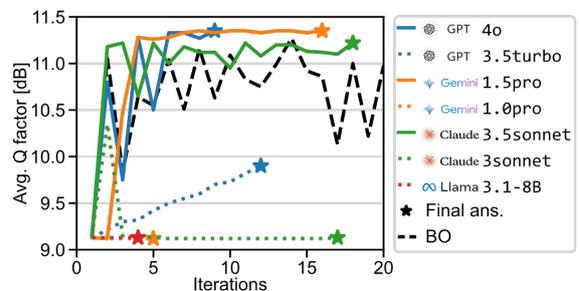

**Fig. 5.** Comparison of OA gain optimization performance.

**Conclusions**

We design and demonstrate the first field trial of an LLM-powered AI Agent for ADON lifecycle management. It autonomously handles multiple representative events in an emulated network lifecycle, including wavelength add/drop and failure handling. We further show that LLM has great potential in tackling intractable tasks for ADON.

**Acknowledgements**
This work was supported by Shanghai Pilot Program for Basic Research - Shanghai Jiao Tong University (21TQ1400213) and National Natural Science Foundation of China (62175145).